\documentclass{amsproc}
\newtheorem{theorem}{Theorem}[section]
\newtheorem{lemma}[theorem]{Lemma}
\theoremstyle{definition}
\newtheorem{definition}[theorem]{Definition}

\theoremstyle{remark}

\numberwithin{equation}{section}

\def\RE{{\mathbb R}}

\def\LS{{L^2_{*}} (\RE^3)}
\def\LT{{L^2_{3}} (\RE^3)}
\def\L{L^2(\RE^3)}
\def\lr{\lim_{r\downarrow 0}}
\def\p{\par\noindent}

\def\HUS{H^{1}_{*}(\RE^3)}
\def\HDS{H^{2}_{*}(\RE^3)}

\def\CH{{\mathcal H}}
\def\nr{{\text{\rm nr}}}
\def\NR{[D(F_m)]_{\nr}}
\def\SY{([D(F_m)]_{\nr}\times [L^2_{*}(\RE^3)]_{\nr}, \Omega_0)}
\def\qed{\hfill\vbox{\hrule\hbox{\vrule\vbox to 7 pt {\vfill\hbox to
         7 pt {\hfill\hfill}\vfill}\vrule}\hrule}\par}

\begin{document}

\title{Delta Interactions and Electrodynamics of Point Particles}

\author{Diego Noja}
\address{Dipartimento di Matematica dell'Universit\`a, Via
Saldini 50, I--20133 Milano, Italy}
\email{noja@berlioz.mat.unimi.it}

\author{Andrea Posilicano}
\address{Dipartimento di Scienze, Universit\`a dell'Insubria, Via
Lucini 3, I--22100 Como, Italy }
\email{posilicano@mat.unimi.it}


\begin{abstract}
We report on some recent work of the authors showing the relations
between singular (point) perturbation of the Laplacian and the
dynamical system describing a charged point particle interacting with
the self--generated radiation field (the Maxwell--Lorentz system) in
the dipole approximation. We show that in the limit of a point
particle, the dynamics of the system is described by an abstract wave
equation containing a selfadjoint operator $H_m$ of the class of point
interactions; the classical Abraham--Lorentz--Dirac third order
equation, or better its integrated second order version, emerges as
the evolution equation of the singular part of the field and is
related to the boundary conditions entering in the definition of the
operator domain of $H_m$. We also give the Hamiltonian structure of the
limit model and, in the case of no external force, we study the reduced
dynamics on the linear stable manifold.
\end{abstract}

\maketitle

\hfill {\em Dedicated to Sergio Albeverio}

\vspace{1cm}

\section{Introduction}
Over a century after the discover of the electron by Thompson and the
first theoretical studies by Lorentz, a satisfying mathematical
description of the interaction of point charged particles and
electromagnetic field is still lacking.  As it is well known, the
Maxwell--Lorentz system, which correctly describes the bulk matter,
looses its meaning in the case of a point particle, due to the
elementary fact that the solutions of the field equations are not
regular enough to be evaluated at a single point, as required by the
particle equation. This classical singularity yields directly to the
need of infinite mass renormalization and to the difficulties which
plague the classical theory of the electron. This theory, as emerges
in the work of Lorentz, Abraham and Dirac among the others, leads to a
reduced equation for the particle alone, the so called Abraham Lorentz
Dirac equation (ALD for short), a third order differential equation
which embodies the interaction between particle and field in an
inertial term (mass renormalization) and a third order term, describing
radiation reaction. The mathematically dubious procedures
involved in its deduction from Maxwell--Lorentz system, and the
presence of unphysical solutions, the so called runaway solutions,
surround the entire subject of a legitimate suspicion, and the usual
way out is to invoke quantum electrodynamics for the solution of the
problem. But, in turn, mathematical foundations of quantum
electrodynamics are far from being clear, and in the opinion of the
present authors and many others a solution to the classical problem is
a first necessary step toward a clarification of the quantum one. In
the following (see [2,3] for complete proofs, more details and references) we
describe the main results obtained in the last years by us in the so
called dipole approximation of classical electrodynamics, a well known
model which in essence amounts to linearize the interaction (see the
Hamiltonian (3.1)) in the Maxwell--Lorentz system. For this model we
show that a limit dynamical system describing the coupled dynamics of
both particle and field indeed exists (theorem 3.3), and is an
abstract wave equation in which the operator part is a point
interaction, the elements of the domain of which have a well precise
relation (see the system (3.5)) with the particle velocity. To
simplify the presentation we introduce first the special case of a
free particle and zero total momentum, where the standard delta
potential appears, and then we give the general case in which an
external force is present. To treat this second case a slight
generalization of the point interaction is needed, where the linear boundary
condition of standard delta interactions are replaced by an affine one in which
the particle momentum enters (see definition 3.2). Such a boundary
condition allows to interpret the solutions of the limit system as
the solutions of the system in which a standard wave equation with a
delta source is coupled with an ordinary second order equation. This
equation is a low order version of the ALD equation.\par 
Our description allows also an immediate Hamiltonian formulation of the
limit dynamics, a result which solves an old problem in classical
electron theory. \par 
In the last paragraph we study, in the free
case, the reduced dynamics on the stable linear manifold, also called,
in the classical literature, the ``non runaway dynamics''. In
particular we show that
the flow of the reduced Hamiltonian system is correlated, through a
canonical transformation, to another Hamiltonian system which lives on
the standard Hilbert space of the free wave equation, the symplectic
form not being however the usual one.
\section{Notations}
\p
\begin{itemize} 
\item[--] $\LS$ is the Hilbert space of square integrable, 
divergence--free, vector fields
on $\RE^3$. \p
\item[--] $M$ is be the projection from $\LT$, the Hilbert space of 
square integrable vector fields on $\RE^3$, onto $\LS$.\p
\item[--] $\langle\cdot,\cdot\rangle $ ( $\| \cdot \|_2$ being the corresponding Hilbert
norm ) denotes the scalar products in 
$\L$, $ \LT$, $\LS$ and also the obvious pairing between an 
element of $\LT$ and one of $\L$ ( the result being a vector in
$\RE^3$ ).\p
\item[--] Given two functions $f$ and $g$ in $\L$, 
$f\otimes g$ is the operator in $\LT$ defined 
by $f\otimes g(A):=f\langle g,A\rangle$. \p
\item[--] $H^s(\RE^3)$, $s\in\RE$, is the usual 
scale of Sobolev--Hilbert spaces, and $H^s_3(\RE^3)$ 
and $H^s_*(\RE^3)$ are defined correspondingly.\p 
\item[--] $\theta$ denotes the Heaviside function.\p 
\item[--] $I(T)$ denotes the compact time interval $[-T,T]$.\p
\item[--] Lip$(\RE^3;\RE^3)$ is the 
space of Lipschitz vector fields.\p 
\item[--] Given a measurable 
non negative function $\rho$, its energy $E(\rho)$ is defined as 
\begin{equation*}
E(\rho):={1\over{4\pi}}\,\int_{\RE^3}\int_{\RE^3}{{\rho(x)\rho(y)}\over{|x-y|}}\,dx\
 dy\ .
\end{equation*}
\end{itemize} 
\section{The Point Limit of the Maxwell--Lorentz System}
Let us consider, on the symplectic vector space $(\HUS\times\LS\times 
\RE^6,\Omega)$, 
$$
\Omega((A_1,E_1,q_1,p_1),(A_2,E_2,q_2,p_2)):=
\langle A_1,E_2\rangle-\langle A_2,E_1\rangle+q_1\cdot p_2-q_2\cdot p_1\ ,
$$
the Hamiltonian associated to the regularized Maxwell--Lorentz 
system in the dipole 
approximation with an external force $F=-\nabla V$, i.e.  
\begin{equation}
\CH_r(A,E,q,p)=2\pi c^2\|E\|^2_2+\frac{1}{8\pi}\,\|\nabla 
A\|^2_2+\frac{1}{2m_r}\,\left| 
p-\frac{e}{c}\,\langle\rho_r,A\rangle\right|^2+V(q)\ ,
\end{equation}
where $c$ is the velocity of light and 
the square integrable density $\rho_r$ describes an extended
particle with electric charge $e$ and radius $r$. 
The corresponding Hamilton equations give rise to the Cauchy problem
\begin{equation}
\cases
&\dot A_r=4\pi c^2E_r\\
&\dot E_r=\frac{1}{4\pi}\,\Delta A_r-{{ e^2}\over m_rc^2}\,M\langle 
\rho_r,
A_r \rangle \rho_r+{{ e}\over {m_rc}}\,Mp_r\rho_r\\
&\dot q_r=\frac{1}{m_r}\,p_r-{e\over m_r c}\,\langle  \rho_r, A_r\rangle\\
&\dot p_r=-\nabla V(q)\\
&A_r(0)=A_0^r\in\HUS,\quad E_r(0)= E_0\in\LS,\\
& q_r(0)=q_0,
\quad p_r(0)=
 p_0\ .
\endcases
\end{equation}
Let us begin with the simplest situation,
i.e. $V={\text {\rm const.}}$ and $p_0=0$. In this case the fields equations decouple from
the particle ones and one is led to study the convergence as $r\downarrow 0$  of 
the
self--adjoint operator
$$
H_r:=-\Delta+\frac{4\pi e^2}{m_rc^2}\,M\cdot\rho_r\otimes\rho_r\ .
$$
This is a first rank perturbation of the Laplacian; its resolvent is
readily calculated and it is given by
$$
(H_{r}+z)^{-1}=(-\Delta+z)^{-1}+\Gamma_r(z)^{-1}
M\cdot (-\Delta+z)^{-1}\rho_r\otimes(-\Delta+z)^{-1}\rho_r\ ,
$$
where Im$\, z\not= 0$,  
\begin{equation*}
\Gamma_r(z)=-{{m_rc^2}\over{4\pi e^2}}-{2\over 3}
\langle(-\Delta+z)^{-1}\rho_r,\rho_r\rangle\ ,
\end{equation*}
and one immediately obtains that, in order to
obtain a non trivial (i.e. different from $-\Delta$) limit, the mass
must be renormalized according to the classical prescription
$$
m_r:=m-{{8\pi e^2}\over{3c^2}}\,E(\rho_r)\ ,
$$
where $m$ is the phenomenological mass. With this definition of $m_r$, 
$(H_{r}+z)^{-1}$  converges in norm to  
$$
R_m(z):=(-\Delta+z)^{-1}+\Gamma_m(z)^{-1}
M\cdot G_{z}\otimes G_{z}\ ,
$$
where
$$
\Gamma_m(\lambda)=-{{mc^2}\over{4\pi e^2}}+{{\sqrt\lambda}\over{6\pi}}\ ,
$$
and
$$
G_{z}(x):={1\over{4\pi}}\,{{e^{- \sqrt z\,|x|}}\over{|x|}}\, ,
\qquad \text{\rm Re}\,\sqrt z\,>0\, .
$$
It is not difficult to show then that $R_m(z)$ is the resolvent of a self--adjoint operator. In more detail one has the
following result, which is no more than an adaptation to our situation
of [1, \S II.1.] as regards the operator aspects and of [4] as regards
the form ones.\p 
\begin{theorem}
As $r\downarrow 0$, i.e as 
$\rho_r(x):=r^{-3}\rho(r^{-1}x)$, $\rho$ a spherically symmetric 
probability density with bounded support, weakly converges 
to $\delta_0$, the self--adjoint operator $H_r$ converges 
in norm resolvent sense in $\LS$ to the self--adjoint operator $H_m$ so defined: \p
1. $ A\in D(H_m)$ if and only if 
$$\exists\, Q_A\in\RE^3\ :\ A_\lambda:= A-\frac{4\pi e}{c }\,MQ_{A}G_\lambda
\in\HDS\, ,\qquad 
-\lambda\in\rho(H_m),\quad \lambda>0\,  ,$$ 
and the 
following boundary condition holds:
$$
\lr\ {1\over{4 \pi r^2}}\,\int_{S_r}\left( A-{{4 \pi e}\over c}\,MQ_{A} 
G_{0}\right) d\mu_{r}=-{{mc}\over e}\,Q_{A}\ ,
$$
where $S_r$ denotes the sphere of radius $r$ and $\mu_r$ is the corresponding 
surface measure.\p
2.
$$
(H_{m}+\lambda)A:=(-\Delta+\lambda)A_{\lambda}\, .
$$
Moreover
$$
\sigma_{ess}(H_m)=\sigma_{ac}(H_m)=[0,+\infty),\quad \sigma_{sc}(H_m)=
\emptyset\ ,
$$
and
$$
\sigma_p(H_m)=\left\{-\left({{3mc^2}\over{2e^2}}\right)^2\right\}
\equiv\{-\lambda_0\}\ ,
$$
where $-\lambda_0$ has a threefold degeneration and  
$$
X^0_j=2\sqrt{2\pi m}\ {c\over e}Me_jG_{{\lambda_0}}\ ,
$$
are the corresponding normalized eigenvectors, where  $\{e_j\}_1^3$ is 
an orthonormal 
basis.\par
If $F_m$ is the quadratic form corresponding to $H_m$ then 
$$
F_m(A,A)+\lambda\|A\|^2_2\\
=\|(-\Delta+\lambda)^{{1\over 2}}A_\lambda\|^2_2+
\left({{4\pi e}\over c}\right)^2\Gamma_m(\lambda)|Q_A|^2\, ,\qquad
\quad\lambda>0\ ,
$$
where
the vector $A$ is  
in the form domain $D(F_m)$ if and only if
$$\exists\,Q_A\in\RE^3\ :\ A_\lambda:=A-{{4\pi e}\over c}\,
MQ_AG_{\lambda}\in \HUS\ .
$$
Finally, given $A\in D(F_m)$, $Q_A$ can be explicitly computed by the formula
$$
Q_A={{3c}\over {2e}}\,\lim_{r\downarrow 0}
r\ {1\over{4\pi r^2}}\,\int_{S_r}A(x)d\mu_r\ .
$$ 
\end{theorem}
\vskip 5pt
The norm resolvent convergence of $H_r$ to $H_m$ implies, by using the
explicit solution for abstract linear wave equations in terms of sine
and cosine operator functions, that if  
$$ \lr\ \|(H_r+\lambda)^{{1\over 2}}\,A_0^r
-(H_m+\lambda)^{{1\over 2}}\,A_0\|_2=0\, ,\qquad\lambda>\lambda_0
$$
(we used the natural distance in the energy norm between elements in the
form domain of $H_r$ and $H_m$), then (see [2, corollary 2.9])
\begin{equation} 
\lr\  \sup_{|t|\le T}\|(H_r+\lambda)^{{1\over 2}}\,A_r(t)
-(H_m+\lambda)^{{1\over 2}}\,A(t)\|_2=0\, .
\end{equation}
Here $A(t)$ is the solution of the Cauchy problem
\begin{equation*}
\cases
&\frac{1}{c^2}\ddot A=-H_m A_m\notag\\
&A(0)=A_0\in D(F_m),\quad \dot A(0)=4\pi c^2 E_0\in\LS
\endcases
\end{equation*} 
This gives the limit field dynamics. \par 
As regards the behaviour of the particle
dynamics in the limit $r\downarrow 0$, by [3, lemma 2.4], relation (3.3) implies 
$$
\lr \ \sup_{|t|\le T}\left|-{e\over{ m_rc}}\langle  \rho_r, 
A_r(t)\rangle-Q_{A(t)}\right|=0\ .
$$
This follows from (here $X\in L_*^2(\RE^3)\,$)
\begin{align}
&-{e\over{ m_rc}}\langle  \rho_r,(H_r+\lambda)^{-1/2}X\rangle
=-{e\over c}{{\langle  \rho_r,(-\Delta+\lambda)^{-{1\over 
2}}X\rangle}\over{m_r}}\notag\\
&-
\frac{2e}{3\pi c  }\int_0^{\infty}\Gamma_r(x+\lambda)^{-1}\langle 
(-\Delta+x+\lambda)^{-1}\rho_r,X\rangle 
{{\langle  (-\Delta+\lambda)^{-1}\rho_r,\rho_r\rangle}\over{m_r}}
\ \frac{dx}{\sqrt 
x}\ ,\notag
\end{align}
$$
Q_{(H_m+\lambda)^{-1/2}X}=\left(\frac{4\pi e}{c}\right)^{-1}
\frac{1}{\pi}
\int_0^{\infty}\Gamma_m(x+\lambda)^{-1}\langle 
G_{x+\lambda},X\rangle \frac{dx}{\sqrt 
x}
$$
and
$$
\lr\ \Gamma_r(\lambda)=\Gamma_m(\lambda)\ ,
$$ 
$$
\lr\ \left|{{\langle  \rho_r,(-\Delta+\lambda)^{-{1\over 2}}
 X\rangle}
\over {m_r}}\right|=0\ ,
$$
$$
\lr\ 
{{\langle  (-\Delta+\lambda)^{-1}\rho_r,\rho_r\rangle}\over{m_r}}=
-{{3c^2}\over{8\pi e^2}}\ .
$$
Therefore
$$
\lr\ \dot q_r=\lr\,-{e\over m_r c}\,\langle  \rho_r, A_r\rangle
=Q_A
$$ 
uniformly in time over compact intervals. \par 
So, to summarize, the dynamics of the system corresponding to the limit of
(3.2) is completely specified (in the case $V={\text{\rm const.}}$, $p_0=0$) by
solving the 
abstract wave equation $c^{-2}\ddot A=-H_m A$ and then recovering the
time evolution of the particle position from the relation $\dot
q=Q_A$. Since the system is linear this can be done explicitly (see
[2, \S3]). \par
Let us now consider the general
situation $V\not={\text{\rm const.}}$, $p_0\not=0$. In this case the fields equations and
the particle ones no more decouple. However, considering the field
equation with an assigned path $t\mapsto p(t)$ not depending on the dynamics,
one is led to study the Cauchy problem  
\begin{equation}
\cases
&\frac{1}{c^2}\ddot A_r=-H_rA_r+{{4\pi e}\over {m_r 
c}}Mp \rho_r\notag\\
&A(0)=A_0^r\in\HUS,\quad \dot A(0)=\dot A_0\in\LS\ .
\endcases
\end{equation}
As in the case $p=0$ one would like to show that the solution of the above system converges to the solution of the Cauchy problem
\begin{equation}
\cases
&\frac{1}{c^2}\ddot A=-H_m^p A\notag\\
&A(0)=A_0\in D(F_m),\quad \dot A(0)=\dot A_0\in\LS\ ,
\endcases
\end{equation}
where $H_m^p$ is a (not necessarily linear) operator to be identified. To this end 
let us consider the system
\begin{equation}
\cases
&\frac{1}{c^2}\ddot A_r=-(H_r+\lambda )A_r+{{4\pi e}\over {m_r 
c}}Mp \rho_r\notag\\
&A(0)=A_0^r\in\HUS,\quad \dot A(0)=\dot A_0\in\LS\ ,
\endcases
\end{equation} 
where the parameter $\lambda>\lambda_0$ is inserted only to make $H_r+\lambda$, 
and successively $H_m+\lambda$, invertible. Its ( mild ) solution is readily 
found and, after an integration by
parts, can be rewritten as 
\begin{align}
A_r(t)=&\cos (ct{(H_r+\lambda)}^{1\over 
2})A_0^r+\sin (ct{(H_r+\lambda)}^{1\over 2}){(H_r+\lambda)}^{-{1\over 2}}
\dot A_0\notag\\
-&{{4\pi e}\over {m_r c}}\int_0^t
\cos (c(t-s){(H_r+\lambda)}^{1\over 2}){(H_r+\lambda)}^{-1} 
M \dot p(s)  \rho_r ds\notag\\
+&{{4\pi e}\over {m_r c}}{(H_r+\lambda)}^{-1}M
p(t)  \rho_r-
{{4\pi e}\over {m_r c}}\cos (ct(H_r+\lambda)^{\frac{1}{2}}) (H_r+\lambda)^{-1}
Mp(0)\rho_r\ .\notag
\end{align}
By [3, lemma 2.5]
$$
\lr\ \left\|
{{4\pi e}\over {m_r c}}{(H_r+\lambda)}^{-{1\over 2}}
M p  \rho_r
+{c\over e}\Gamma_{m}(\lambda)^{-1}{(H_m+\lambda)}^{1\over 2}
Mp G_\lambda\right\|_2=0
$$ 
and so,
by thm. 3.1, if 
$$
\lr\ \|(H_r+\lambda)^{{1\over 2}}A_0^r
-(H_m+\lambda)^{{1\over 2}}A_0\|_2=0\ ,
$$
then
$$
\lr\ \sup_{|t|\le T}\|(H_r+\lambda)^{{1\over 2}}A_r(t)
-(H_m+\lambda)^{{1\over 2}}A(t)\|_2=0\ ,
$$
where
\begin{align}
A(t)=&\cos (ct{(H_m+\lambda)}^{1\over 
2})A_0+\sin (ct{(H_m+\lambda )}^{1\over 2}){(H_m+\lambda )}^{-{1\over 2}}\dot 
A_0\notag\\
+&{c\over e}\Gamma_{m}(\lambda)^{-1}\int_0^t \cos (c(t-s){(H_m+\lambda)}^{1\over 2})
M\dot p(s) G_\lambda\ ds\notag\\
-&{c\over e}\Gamma_{m}(\lambda)^{-1}Mp(t)G_\lambda
+{c\over e}\Gamma_{m}(\lambda)^{-1}\cos (ct(H_m+\lambda)^{\frac{1}{2}})
Mp(0)G_\lambda\ .\notag
\end{align}
Finally, again integrating by parts, it is easily seen that 
$$A_p(t):=A(t)+{c\over e}\Gamma_{m}(\lambda)^{-1}Mp(t)G_\lambda$$
solves the Cauchy problem 
\begin{equation}
\cases
&\frac{1}{c^2}\ddot A_p=-(H_m+\lambda )A_p
+{c\over e}\Gamma_{m}(\lambda)^{-1}M\ddot pG_\lambda
\notag\\
&A_p(0)= A_0+{c\over e}\Gamma_{m}(\lambda)^{-1}Mp(0)G_\lambda,
\quad \dot A_p(0)=\dot A_0+{c\over e}\Gamma_{m}(\lambda)^{-1}M\dot p(0)G_\lambda\ ,
\endcases
\end{equation} 
and so $A(t)$ solves the Cauchy problem
\begin{equation}
\cases
&\frac{1}{c^2}\ddot A=-(H_m+\lambda )
\left(A+{c\over e}\Gamma_{m}(\lambda)^{-1}M pG_\lambda\right)
\notag\\
&A(0)= A_0\in D(F_m),\quad \dot A(0)=\dot A_0\in \LS\ .
\endcases
\end{equation}
This induces us to define, on the domain 
$$
D(H_m^p):=D(H_m)-{c\over e}\Gamma _m(\lambda)^{-1}
MpG_{\lambda}\ ,
$$
the affine operator $H_m^p$ according to 
$$
(H_m^p+\lambda )A:=(H_m+\lambda)
\left(A+{c\over e}\Gamma_{m}(\lambda)^{-1} 
MpG_\lambda\right) \, .
$$
Alternatively (see [3, lemma 4.1]) $H_m^p$ can be defined in the
following way:\p  
\begin{definition} $\quad$ 1. $A\in D(H_{m}^p)$ if and only if 
$$\exists\, Q_A\in\RE^3\ :\ A_\lambda:= A-\frac{4\pi e}{c }\,MQ_{A}G_\lambda
\in\HDS\, ,\qquad
-\lambda\in\rho(H_m),\quad \lambda>0\ ,$$ 
 and the 
following boundary condition holds:
$$
\lr\ {1\over{4 \pi r^2}}\,\int_{S_r}\left( A-{{4 \pi e}\over c}\,MQ_{A} 
G_{0}\right) d\mu_{r}=-{{mc}\over e}\,Q_{A}+ {c\over e}\,p\ .
$$
2.
$$
(H_{m}^p+\lambda)A:=(-\Delta+\lambda)A_{\lambda}\, .$$
\end{definition} 
\vskip 5pt
Note that, as it is evident, the affine operator $H^p_m$ reduces, for $p=0$, to the linear operator $H_m$ describing the standard point interaction.\par 
The same considerations leading to the definition of $H_m^p$ give,
coupled with a fixed point argument and estimates uniform in $r$, the following
\begin{theorem} {\rm (see [3, thm. 3.4])} Let $V$ such that $\nabla V\in 
\text{\rm Lip}(\RE^3;\RE^3)$, 
$|\nabla V(x)|\le K\,(1+|x|)$, $\lambda>\lambda_0$, and 
$E_0\in\LS$. Let $A_0^r\in \HUS$, $A_0\in 
D(F_m)$, such that 
\begin{equation}
\lr\ \|(H_r+\lambda)^{{1\over 2}}A_0^r
-(H_m+\lambda)^{{1\over 2}}A_0\|_2=0\ .
\end{equation}
Then there exists $T>0$, not depending on $r$, such that, denoting by 
$$(A_r,E_r,q_r,p_r)\in C(I(T);\HUS)\times C(I(T);\LS)
\times C^2(I(T);\RE^3)\times C^1(I(T);\RE^3)$$ 
the unique mild solution of the Cauchy problem (3.2), one has
\begin{align}
&\lr\ \sup_{|t|\le T}\|(H_r+\lambda)^{{1\over 2}}A_r(t)
-(H_m+\lambda)^{{1\over 2}}A(t)\|_2=0,\notag\\
&\lr\ \sup_{|t|\le T}\| E_r(t)- E(t)\|_2=0,\notag\\ 
&\lr\ \sup_{|t|\le T}|\dot q_r(t)-\dot q(t)|+
\sup_{|t|\le T}| q_r(t)- q(t)|=0,\notag\\
&\lr\ \sup_{|t|\le T}|\dot p_r(t)-\dot p(t)|+
\sup_{|t|\le T}| p_r(t)- p(t)|=0\ ,\notag
\end{align}
where $$(A,E,q,p)\in C(I(T);D(F_m))\times  
C(I(T);\LS)\times C^1(I(T);\RE^3)\times C^1(I(T);\RE^3)$$
denotes the unique mild solution of the Cauchy problem
\begin{equation}
\cases
&\dot A=4\pi c^2E\\
&\dot E=-\frac{1}{4\pi}\,H_{m}^pA\\
&\dot q=Q_A\\
&\dot p=-\nabla V(q)\\
&A(0)=A_0\in D(F_m),\quad E(0)=E_0\in \LS,\\
&q(0)=q_0,\quad p(0)=p_0\ .
\endcases
\end{equation}
\end{theorem}
\vskip 5pt
An alternative description of the limit dynamics defined by the previous 
system is
provided by the following
\begin{theorem} {\rm (see [3, thm. 4.2])} Given $V$ such that $\nabla V\in\text{\rm Lip}(\RE^3;\RE^3)$, 
let $$(A,E,q,p)\in C^1([0,T];D(F_m))\times  
C^1([0,T];\LS)\times C^2([0,T];\RE^3)\times C^1([0,T];\RE^3)\, ,
$$ 
$A_0\in D(H_m^{p_0})$, $E_0\in D(F_m)$, be the unique strict solution of the Cauchy 
problem (3.5). Then $$
A(t)=A_f(t)+\frac{4\pi e}{c}\,MA_\delta(t)\ ,
$$
where $A_f(t)$ is the solution of the free wave equation with initial 
data $A_0, E_0$ and $A_\delta$ is the retarded potential of the
source $Q_A\delta_0$, i.e.
$$
A_\delta(t,x)=\frac{1}{4\pi}\,\frac{\theta(ct-|x|)}{|x|}Q_{A(t-|x|/c)}\ .
$$
Moreover $Q_A$ satisfies 
the equation
\begin{equation}
\dot Q_{A(t)}=c\sqrt{\lambda_0}Q_{A(t)}+\frac{3c^2}{2e}\,A_f(t,0)
-\frac{3c^3}{2e^2}\,p(t)\ .
\end{equation}
\end{theorem}
\vskip 5pt
Let us sketch the proof of the above theorem. Given an arbitrary function $Q(t)$, 
consider the function  
$$
A(t,x):=A_f(t,x)+\frac{1}{4\pi}\,\frac{\theta(ct-|x|)}{|x|}Q{(t-|x|/c)}\ .
$$
It solves the distributional equation
$$
{1\over{c^2}}\ddot A=\Delta A+{{4\pi e}\over c}MQ\delta_0\ .
$$
Kirchhoff formula shows that $A_f$ gives no contribution to $Q_A$ 
(see [1, \S3]) and so $Q_{A(t)}=Q(t)$.
Moreover, by [3, lemma 4.1] and [2, thm. 3.3] there follows 
$$A-\frac{4\pi e}{c}MQG_\lambda\in 
H_*^2(\RE^3)\ .
$$ 
 Therefore 
if $A(t)\in D\left(H_m^{p(t)}\right)$, since
$$
{1\over{c^2}}\ddot A=\Delta A+\frac{4\pi e}{c }MQ\delta_0=
\Delta\left(A-\frac{4\pi e}{c}MQG_{\lambda}\right)
+\lambda\frac{4\pi e}{c}MQG_{\lambda}=-H_{m}^pA\ ,
$$
then the thesis will follow from unicity of the solution of (3.5).
The conditions on $Q(t)$ leading to $A(t)\in D\left(H_m^{p(t)}\right)$ are found 
as follows. By an elementary integration
\begin{align}
\lr& {1\over{4 \pi r^2}}\int_{S_r}\left( A-{{4 \pi e}\over c}MQ 
G_{0}\right) d\mu_{r} (x)\notag\\
=&A_f(t,0)+\frac{2}{3}\frac{e}{c^2}\lr\left(\frac{Q(t-r/c)-Q(t)}{r/c}\right)
\notag\\
=& A_f(t,0)-\frac{2}{3}\frac{e}{c^2}\dot Q(t)\notag
\end{align}
and so $A$ satisfies the boundary condition in Definition 3.2 if and only 
if $Q(t)$ solves (3.6). \par 
Let us remark that the above theorem gives the connection with the
traditional description: indeed the field variable satisfies a standard wave
equation with a point source and the equation satisfied by $Q_A$
(i.e. $\dot q$) is the integrated version of the ALD equation
$$
m\ddot q(t)=\frac{2e^2}{3c^3}\dddot q(t)
-{e\over c}\dot A_f(t,0) + F(q(t))\ .
$$
Moreover the above argument also gives the converse statement:\p
the solution of the distributional Cauchy problem
\begin{equation*}
\cases
&{1\over{c^2}}\ddot A=\Delta A+\frac{4\pi e}{c }M\dot q\delta_0\\
&\ddot q(t)=c\sqrt{\lambda_0}\dot q(t)+\frac{3c^2}{2e}\,A_f(t,0)
-\frac{3c^3}{2e^2}\,p(t)\\
&\dot p=-\nabla V(q)\\
&A(0)=A_0\in D(H^{p_0}_m),\quad \dot A(0)=4\pi c^2E_0\in D(F_m),\\
&q(0)=q_0,\quad \dot q(0)=Q_{A_0},\quad p(0)=p_0
\endcases
\end{equation*}
solves (3.5).
Also note that the equivalence between the two descriptions holds true if
and only if the initial data for the field are chosen coherently with
the particle's ones (i.e. $\dot q_0=Q_{A_0}$ and $\ddot q_0=Q_{\dot A_0}$).\par
Let us now come to the Hamiltonian character of system (3.5). By
definition 3.2, and by $D(H_{m}^p)\subset D(F_m)$, one can check that
$$
\langle 
H_{m}^pA_1,A_2\rangle=F_m(A_1,A_2)+4\pi p\cdot Q_{A_2}\ .
$$ 
Therefore equations (3.5) 
are nothing but the Hamilton equations corresponding to the 
(degenerate) Hamiltonian
$$
\CH_m(A,E,q,p):=2\pi c^2\|E\|^2_2+\frac{1}{8\pi}\,F_m(A,A)+p\cdot 
Q_A+V(q)\ ;
$$ 
this is defined on the symplectic vector space $(D(F_m)\times\LS\times 
\RE^6,\Omega)$.
Moreover one has the following convergence result:
\begin{theorem} {\rm (see [3, thm.4.5])} Let
$$
\CH_r(A,E,q,p)=2\pi c^2\|E\|^2_2+\frac{1}{8\pi}\,\|\nabla 
A\|^2_2+\frac{1}{2m_r}\,\left| 
p-\frac{e}{c}\,\langle\rho_r,A\rangle\right|^2+V(q)
$$
be the Hamiltonian giving the equations (3.2), let $E\in\LS$, 
$(q,p)\in\RE^6$, and let $A_r\in\HUS$, $A\in D(F_m)$ satisfy the 
condition (3.4). Then  
$$
\lr\ 
\CH_r(A_r,E,q,p)=
\CH_m(A,E,q,p)\ .
$$
\end{theorem}

\section{The Non Runaway Dynamics}
The negative eigenvalue in the spectrum of the operator
$H_m$ gives rise to unstable behaviour which corresponds, in classical
electron theory, to the presence of the so called ``runaway solutions''. 
In this section
we briefly describe, in the free case and, for simplicity of
presentation, vanishing particle momentum, the reduced dynamics on the
stable manifold.\par Given any vector subspace ${\mathcal
V}\subseteq\LS$, we define the corresponding ``non runaway'' subspace
$[{\mathcal V}]_{\nr}$ by
$$
[{\mathcal V}]_{\nr}:=
\{\ A\in {\mathcal V}\ :\ \langle A,X_j^0\rangle=0,\ j=1,2,3\ \}\ ,
$$
$X^0_j$ being the eigenvectors corresponding to $-\lambda_0$ (see thm. 3.1).
Observe that, if $A=A_\lambda+\frac{4\pi e}{c}\,MQ_AG_{\lambda}
\in D(F_m)$, then one has 
\begin{align}
\langle A,X_j^0\rangle=&
\langle A_{\lambda},X_j^0\rangle+\frac{4\pi e}{c}\,
\langle MQ_AG_\lambda,X_j^0\rangle\notag\\
=&2\sqrt{2\pi m}\frac{c}{e}\,\left(\langle 
A^j_\lambda,G_{\lambda_0}\rangle+
\frac{8\pi e}{3c}\,
Q_A^j\langle G_\lambda,G_{\lambda_0}\rangle\right)\ .\notag
\end{align}
Therefore $A\in [D(F_m)]_{\nr}$ if and only if
$$
Q_A=-\frac{3c}{8\pi e}\,\frac{\langle G_{\lambda_0},A_\lambda
\rangle}{\langle G_\lambda,G_{\lambda_0}\rangle}\ ,
$$
so that any $A\in [D(F_m)]_{\nr}$ is 
univocally determined by 
its regular part $A_\lambda$. This implies, since 
$\langle 
G_{\lambda_0},G_{\lambda_0}\rangle^{-1}=8\pi\sqrt{\lambda_0}$, 
that the map 
$$
\Phi:\HUS\to [D(F_m)]_{\nr},\quad 
\Phi A:=
A+ M P_A G_{\lambda_0},\quad 
P_A:=-12\pi\sqrt{\lambda_0}\langle G_{\lambda_0},A\rangle\ ,
$$
is bijective, with inverse given by 
$$
\Phi^{-1}:[D(F_m)]_{\nr}\to\HUS,\quad 
\Phi^{-1}A:=A-\frac{4\pi e}{c}MQ_AG_{\lambda_0}\equiv A_{\lambda_0}\ .
$$ 
\begin{lemma} $\Phi$ 
is a continuous bijection between $\HUS$ and $\NR$. Moreover 
\begin{equation}
\Phi^{-1}([D(H_m)]_{\nr})=\{A\in\HDS\ :\ A(0)=0\}\ .
\end{equation}
\end{lemma}
{\it Proof.} If $\lambda>\lambda_0$ we can use $\|A\|_\lambda:=
F_m^\lambda(A,A)$ as a norm on $\NR$. Then 
\begin{align}
&\|\Phi A\|_{\lambda}^2=\|A+ MP_A(G_{\lambda_0}-G_{\lambda})+ MP_A
G_\lambda\|_{\lambda}\notag\\
=&\|(-\Delta+\lambda)^{\frac{1}{2}}\,(A+ 
MP_A(G_{\lambda_0}-G_{\lambda}))\|^2_2+\frac{2}{3}\,\Gamma_m(\lambda)|P_A|^2\notag\\
\le& 2(\|(-\Delta+\lambda)^{\frac{1}{2}}A\|^2_2
+\|(-\Delta+\lambda)^{\frac{1}{2}}MP_A(G_{\lambda_0}-G_{\lambda})\|^2_2)
+\frac{2}{3}\,\Gamma_m(\lambda)|P_A|^2\notag\\
\le & 2\|(-\Delta+\lambda)^{\frac{1}{2}}A\|^2_{2}
+c_1|P_A|^2\le 2\|(-\Delta+\lambda)^{\frac{1}{2}}A\|^2_{2}+c_2\|A\|^2_{2}\notag\\
\le & c_3\|(-\Delta+\lambda)^{\frac{1}{2}}A\|^2_{2}\ .\notag
\end{align}
Since, for any $\lambda\not=\lambda_0$, 
$A=A_\lambda+\Gamma_m(\lambda)^{-1}MA_\lambda(0)G_\lambda\in [D(H_m)]_{\nr}$
if and only if $A_\lambda(0)=\Gamma_m(\lambda)P_{A_\lambda}$, one has 
\begin{align}
(\Phi^{-1}A)(0)=& A_\lambda(0)+ (MP_A(G_{\lambda}-G_{\lambda_0}))(0)\notag\\
=&A_\lambda(0)+ 
\frac{2}{3}\frac{\sqrt{\lambda_0}-\sqrt\lambda}{4\pi}P_A\notag\\
=&A_\lambda(0)-\Gamma_m(\lambda)P_{A_\lambda}=0
\ .\notag
\end{align}
Then $\Phi^{-1}$ is continuous by bijectivity and by the open mapping 
theorem.\qed 
\vskip 3pt\p Obviously 
the map $\Phi$ can be extended to the whole $\LS$, 
giving rise the orthogonal projection onto 
$ [L^2_{*}(\RE^3)]_{\nr}$. \vskip 3pt\p
Let us now consider the symplectic space $\SY$, where 
$\Omega_0$ denotes the canonical 
symplectic form induced by $\langle\cdot ,\cdot \rangle$, i.e. 
$$\Omega_0((A_1,E_1),(A_2,E_2) )=
\langle A_1,E_2\rangle-\langle A_2,E_1\rangle\ .
$$
On $\SY$ we 
have the {\it non negative} Hamiltonian 
\begin{align}
&{\mathcal H}^{+}_{m}(A,E)=2\pi c^2\,\| E\|^2_2+\frac{1}{8\pi}\,F_m(A,A)\notag\\
=&2\pi c^2\,\| E\|^2_2+\frac{1}{8\pi}\,
\left(\|\nabla A_{\lambda_0}\|^2_2+\frac{\sqrt{\lambda_0}}{12\pi}
\left(\frac{4\pi e}{c}\right)^2|Q_A|^2\right)
\ ,\notag
\end{align}
with corresponding Hamiltonian vector field 
$$
X_{{\mathcal H}^+_m}(A,E)=\left(4\pi c^2 E,-\frac{1}{4\pi}H_m A\right)\ ,
$$ 
defined on the domain $[D(H_m)]_{\nr}\times \NR$. If we pull--back
$\Omega_0$ and ${\mathcal H}^+_m$ to $\HUS\times\LS$ by using the map
$$\Psi:=\Phi\times \Phi:\HUS\times \LS\to [D(F_m)]_\nr\times[\LS]_\nr\ ,$$
we obtain the following\p 
\begin{theorem}  
$\quad$ 1. $(\HUS\times \LS,\Omega_{\nr})$ is a symplectic 
space, where the (weakly) nondegenerate symplectic form $\Omega_{\nr}$ 
is given by
\begin{align}
&\Omega_{\nr}((A_1,E_1),(A_2,E_2)):=\Psi^*
\Omega_{0}((A_1,E_1),(A_2,E_2))\notag\\
=&\Omega_{0}(( A_1,E_1),( A_2,E_2))-\frac{1}{12\pi\sqrt{\lambda_0}}\,
\left(P_{A_1}\cdot P_{E_2}-P_{A_2}\cdot P_{E_1}\right)
\ .\notag
\end{align}
2. Defining 
$$
X_{\CH_{\nr}}:\HDS\times\HUS\to\HUS\times\LS
$$
$$
X_{\CH_{\nr}}(A,E):=\left(4\pi c^2 E,\frac{1}{4\pi}\,\Delta
A+\frac{3}{2}\,\sqrt{\lambda_0}MA(0) G_{\lambda_0}\right)\ ,
$$
one has, for any $(A,E)\in\HDS\times\HUS$, 
$$\frac{1}{2}\,\Omega_{\nr}(X_{\CH_{\nr}}(A,E),(A,E))=\CH_{\nr}(A,E)\ ,
$$
where
$$
{\mathcal H_{\nr}}:\HUS\times\LS\to\RE
$$ 
\begin{align}&
{\mathcal H}_{\nr}(A,E):=\Psi^*{\mathcal H}^+_{m}(A,E)
\notag\\
=&
2\pi c^2\,\| E\|^2_2+\frac{1}{8\pi}\,\|\nabla A\|_2^2
-\frac{c^2}{6\sqrt{\lambda_0}}\,|P_{E}|^2
+\frac{\sqrt{\lambda_0}}{96\pi^2}\,|P_A|^2
\ .
\notag
\end{align}
i.e. $X_{\CH_{\nr}}$ is the (unique) Hamiltonian vector field 
corresponding to $\CH_{\nr}$.\p
3. The Hamiltonian vector fields $X_{\CH^+_m}$
and $X_{\CH_{\nr}}$ are $\Psi$--correlated, i.e. 
\begin{equation}
X_{\CH^+_{m}}\circ\Psi=\Psi\circ X_{\CH_{\nr}}\ ,
\end{equation}
and
\begin{equation}
U_m(t)\circ\Psi=\Psi\circ U_{\nr}(t)\ ,
\end{equation}
where $U_m(t)$ and $U_{\nr}(t)$, $t\in\RE$, denote the one parameter
groups of canonical transformation of $([D(F_m)]_\nr\times
[\LS]_\nr,\Omega_0)$ and 
$(\HUS\times\LS,\Omega_{\nr})$
given by the flows of $X_{{\mathcal H}^+_m}$ and $X_{\CH_{\nr}}$
respectively. Moreover, if $U_f(t)$, $t\in\RE$, denotes the flow, on
$\HUS\times\LS$, given by solving the free wave equation, then
\begin{equation}
U_m(t)_{\left|\,[D(H_m)]_{\nr}\times [D(F_m)]_{\nr}\right.}=\Psi\circ U_f(t)\circ\Psi^{-1}\ 
.
\end{equation}
\end{theorem}
{\it Proof.} 1. A simple calculation shows that
\begin{align}
&\Omega_{0}((\Phi A_1,\Phi E_1),(\Phi A_2,\Phi E_2))=
\langle\Phi A_1,\Phi E_2\rangle-\langle\Phi A_2,\Phi E_1\rangle\notag\\
=&\Omega_{0}(( A_1,E_1),( A_2,E_2))-\frac{1}{12\pi\sqrt{\lambda_0}}\,
\left(P_{A_1}\cdot P_{E_2}-P_{A_2}\cdot P_{E_1}\right)\ ,\notag
\end{align}
\p
2. Since
\begin{align}
&F_{m}(\Phi A,\Phi A)=
\|(-\Delta+\lambda_0)^{{1\over 2}}A\|_2^2-\lambda_0\|\Phi A\|^2_2\notag\\
=&\|(-\Delta+\lambda_0)^{{1\over 2}}A\|_2^2-
\lambda_0\langle A,\Phi A\rangle\notag\\
=&\|\nabla A\|^2_2+\frac{\sqrt{\lambda_0}}{12\pi}\, |P_A|^2\ ,\notag
\end{align}
$$
\|\Phi E\|_2^2=\langle E,\Phi E\rangle=
\| E\|^2_2-\frac{1}{12\pi\sqrt{\lambda_0}}\, |P_{E}|^2\ ,
$$
posing $4\pi H_{\nr}A=-\Delta A-6\pi\sqrt{\lambda_0}MA(0) 
G_{\lambda_0}$, we need to verify the relation
$$
-\langle\Delta A,A\rangle+\frac{\sqrt{\lambda_0}}{12\pi}\,|P_A|^2=4\pi 
\langle H_{\nr} 
A,A\rangle-\frac{1}{12\pi\sqrt{\lambda_0}}\,P_{4\pi H_{\nr}A}\cdot P_A\ .
$$
Since
\begin{align}
P_{\Delta A}=&
-12\pi\sqrt{\lambda_0}(-\langle G_{\lambda_0},({-\Delta}+\lambda_0)A\rangle
+\lambda_0\langle G_{\lambda_0},{ A}\rangle)\notag\\
=& 12\pi\sqrt{\lambda_0}A(0)+\lambda_0P_A\ ,\notag
\end{align}
and
$$
P_{M A(0)G_{\lambda_0}}=-12\pi\sqrt{\lambda_0}\frac{2}{3}\,
\frac{1}{8\pi\sqrt{\lambda_0}}\,A(0)=-A(0)\ ,
$$
one has
\begin{align}
&4\pi \langle H_{\nr} 
A,A\rangle-\frac{1}{12\pi\sqrt{\lambda_0}}\,P_{4\pi H_{\nr}A}\cdot P_A\notag\\
=&-\langle\Delta A,A\rangle-\frac{1}{2}\,A(0)\cdot P_A\notag\\
+&\frac{1}{12\pi\sqrt{\lambda_0}}\,\left(12\pi\sqrt{\lambda_0}A(0)+\lambda_0P_A
\right)\cdot P_A-\frac{1}{2}\,A(0)\cdot P_A\notag\\
=&-\langle\Delta A,A\rangle+\frac{\sqrt{\lambda_0}}{12\pi}\,|P_A|^2\ .\notag
\end{align}
3. $X_{\CH_{\nr}}$ generates a group of continuous (w.r.t. the 
Hilbert norm on $\HUS\oplus\LS$) linear transformations since we can 
write 
$$4\pi H_{\nr}+\lambda_0 =
(1-6\pi\sqrt{\lambda_0}G_{\lambda_0}\otimes G_{\lambda_0})
\circ (-\Delta+\lambda_0 )\ .$$
Moreover such transformations are symplectic, since 
$X_{\CH_{\nr}}$ is Hamiltonian. Formula $(4.2)$ follows from the definitions of 
$\Omega_{\nr}$ and $\CH_{nr}$, and $(4.3)$ follows from (4.2) by the unicity of 
generators. Finally, $(4.4)$ follows from (4.1) and the definition of $X_{\CH_{nr}}$.
${\qquad}$\qed

\end{document}